\documentclass[conference]{IEEEtran}%
\usepackage{amssymb}
\usepackage{graphicx}
\usepackage{amsmath}
\usepackage{amsbsy}
\usepackage{amsfonts}
\usepackage{amssymb,dsfont,physics}
\usepackage{tikz}
\usetikzlibrary{shapes,snakes}
\usepackage{accents}%
\providecommand{\U}[1]{\protect\rule{.1in}{.1in}}

\newcommand{\SP}[1]{{{\textcolor{black}{#1}}}}

\begin{document}

\title{Energetic Considerations in \\Quantum Target Ranging}

\author{\IEEEauthorblockN{Athena Karsa$^{\star}$ and Stefano Pirandola}
\IEEEauthorblockA{Department of Computer Science\\ University of York, York, YO10 5GH, UK\\
$^{\star}$Email: ak1674@york.ac.uk}}


\date{\today}

\maketitle

\begin{abstract}
While quantum illumination (QI) can offer a quantum-enhancement in target detection, its potential for performing target ranging remains unclear. With its capabilities hinging on a joint-measurement between a returning signal and its retained idler, an unknown return time makes a QI-based protocol difficult to realise. This paper outlines a potential QI-based approach to quantum target ranging based on recent developments in multiple quantum hypothesis testing and quantum-enhanced channel position finding (CPF). Applying CPF to time bins, one finds an upper-bound on the error probability for quantum target ranging. However, using energetic considerations, we show that for such a scheme a quantum advantage may not
\SP{be proven with current mathematical tools.}
\end{abstract}



\section{Introduction}

Quantum illumination (QI)~\cite{pirandola2018advances,lloyd2008enhanced,tan2008quantum,karsa2020generic} is an entanglement-based protocol which, using an optimum quantum receiver, promises a 6 dB advantage in detecting a low-reflectivity object embedded in a bright thermal background. This advantage is present when a very low energy source is used and despite the fact that over the course of the protocol all entanglement is lost~\cite{zhangexp,lopaevaexp,zhangent,barzanjehmicrowave,sacchi2005entanglement}. Since its inception several prototype experiments have been demonstrated~\cite{shabirQI,luong2019receiver} however there are still many aspects inhibiting quantum radar's readiness for real-world implementation~\cite{pirandola2018advances,brandsema2018current}; one of which, also the subject of this paper, is that of quantum target ranging. The issue with target ranging, of course, is the fact that signal-idler recombination becomes problematic. These two modes' return at the receiver must be synchronised to ensure that the joint measurement procedure is a success. 

Quantum-enhanced channel position finding (CPF)~\cite{zhuang2020entanglement} could form a crucial component to extending the QI protocol from simple detection to actual measurement. Fundamentally a pattern recognition problem, CPF is based on multiple quantum hypothesis testing~\cite{helstrom1969quantum,cheflesQSD,barnettQSD} and quantum channel discrimination~\cite{hayashi2017quantum} where the goal is to locate the `target' channel amongst an ensemble of `background', or reference, channels~\cite{zhuang2020b}. The problem may be reformulated for target metrology by parameterising the quantum channels under study by some measurable target property, such as its position, range and velocity.

This paper investigates a quantum-enhanced target ranging protocol based on CPF and QI and studies it as a potential solution to the QI ranging problem. In section~\ref{secCPF} we outline the CPF problem and provide the relevant bounds on error probability in the cases where classical and entangled light sources are employed. Further, we briefly review benefits afforded by the use of a conditional-nulling (CN) receiver~\cite{zhuang2020entanglement} for the entangled case. In section~\ref{secQTR} we apply the CPF \SP{model} 
to target ranging and determine whether or not, under any physical parameter constraints, a quantum-enhancement in ranging can be realised using such a multi-array QI-based approach.

\section{Quantum channel position finding}\label{secCPF}

\begin{figure}[t]
\resizebox{\linewidth}{!}{\begin{centering}
\begin{tikzpicture}
\node[draw,thick] at (0,1) {$\mathcal{B}_1$};
\node[draw,thick] at (0,0) {$\mathcal{B}_2$};
\filldraw [black] (0,-0.75) circle (0.75pt);
\filldraw [black] (0,-1) circle (0.75pt);
\filldraw [black] (0,-1.25) circle (0.75pt);
\node[draw,thick] at (0,-2) {$\mathcal{B}_{m}$};
\draw[thick,->,red] (-2,1.1) -- (-0.5,1.1);
\draw[thick,->,red] (-2,0.1) -- (-0.5,0.1);
\draw[thick,->,red] (-2,-1.9) -- (-0.5,-1.9);
\draw[thick,->,red] (0.5,1.1) -- (2,1.1);
\draw[thick,->,red] (0.5,0.1) -- (2,0.1);
\draw[thick,->,red] (0.5,-1.9) -- (2,-1.9);
\draw[thick,->] (-2,0.9) -- (-0.5,0.9);
\draw[thick,->] (-2,-0.1) -- (-0.5,-0.1);
\draw[thick,->] (-2,-2.1) -- (-0.5,-2.1);
\draw[thick,->] (0.5,0.9) -- (2,0.9);
\draw[thick,->] (0.5,-0.1) -- (2,-0.1);
\draw[thick,->] (0.5,-2.1) -- (2,-2.1);
\node at (-2.25,1.15) {$\psi_{SI}$};
\node at (-2.25,0.15) {$\psi_{SI}$};
\node at (-2.25,-1.85) {$\psi_{SI}$};
\node at (-2.25,0.85) {$\sigma$};
\node at (-2.25,-0.15) {$\sigma$};
\node at (-2.25,-2.15) {$\sigma$};
\begin{scope}
\clip (2.2,0.75) rectangle (2.45,1.25);
\draw[thick,fill=black!60] (2.2,1) circle(0.25);
\draw[thick] (2.2,0.75) -- (2.2,1.25);
\end{scope}
\begin{scope}
\clip (2.2,-0.25) rectangle (2.45,0.25);
\draw[thick,fill=black!60] (2.2,0) circle(0.25);
\draw[thick] (2.2,-0.25) -- (2.2,0.25);
\end{scope}
\begin{scope}
\clip (2.2,-1.75) rectangle (2.45,-2.25);
\draw[thick,fill=black!60] (2.2,-2) circle(0.25);
\draw[thick] (-2.2,1.75) -- (-2.2,2.25);
\end{scope}
\draw[thick,dashed] (2.5,1) -- (3,1);
\node at (3.3,1) {$\mu_1$};
\draw[thick,dashed] (2.5,0) -- (3,0);
\node at (3.3,0) {$\mu_2$};
\draw[thick,dashed] (2.5,-2) -- (3,-2);
\node at (3.3,-2) {$\mu_m$};
\draw[ dashed,thick, blue] (-1.8,1.1) -- (-1.8,1.5) -- (1.8,1.5) -- (1.8,1.1);
\draw[ dashed,thick, blue] (-1.8,0.1) -- (-1.8,0.5) -- (1.8,0.5) -- (1.8,0.1);
\draw[dashed,thick, blue] (-1.8,-1.9) -- (-1.8,-1.5) -- (1.8,-1.5) -- (1.8,-1.9);
\draw[thick] (3.4,1.5) -- (3.7,1.5) -- (3.7,-2.4) -- (3.4,-2.4);
\draw[thick,dashed] (3.7,0) -- (4,0);
\node at (4.2,0) {$\tilde{\mu}$};
\end{tikzpicture}
\end{centering}}
\caption{Schematic diagram of channel position finding (CPF). For $i=1,\dots, m$ we have an ensemble of boxes $B_i$ and the task is locate the single target channel amongst background channels. In the classical strategy, our source $\sigma$ is sent through the channel (black) with the output going straight to the receiver for post-processing. In the quantum strategy, we consider a maximally-entangled two-mode source $\psi_{SI}$ comprising a signal (red), sent through the channel, and an idler (blue) which recombines with the output at the receiver. For each strategy, outputs from each box are combined to yield a final result, $\tilde{\mu}$, giving the target's location with some error probability.}
\label{diagram}
\end{figure}
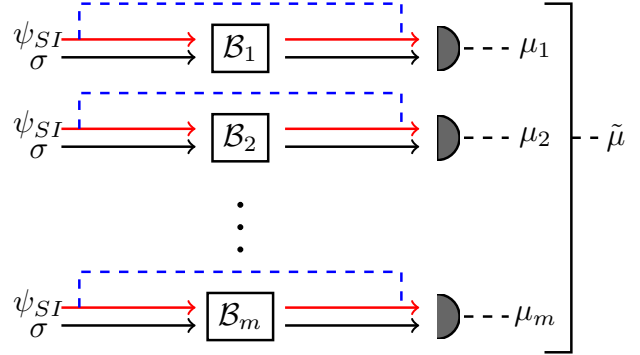

Consider the task of quantum CPF comprising $m \geq 2$ input-output black boxes as shown in Fig.~\ref{diagram}. For $i=1,2,\dots m$, the $i$th box $\mathcal{B}_i$ contains either a background channel $\mathcal{E}^{(B)}$ or some target channel $\mathcal{E}^{(T)} \neq \mathcal{E}^{(B)}$ and the task is to locate its position. The assumption is made that the target channel occupies one box only, i.e., joint probabilities of the form $\mathrm{P}(\mathcal{B}_i=\mathcal{E}^{(T)},\mathcal{B}_j = \mathcal{E}^{(T)})$ are all zero, and the target channel is in one of the boxes with certainty: $\mathrm{P}(\mathcal{B}_i =\mathcal{E}^{(B)}\,\, \forall i)=0$. Identification of the target channel is a problem of symmetric quantum hypothesis testing (in which both error types are given equal weight and minimised simultaneously) where the task is to discriminate between $N$ hypotheses given by
\begin{equation}
H_i : \mathcal{B}_i = \mathcal{E}^{(T)} \quad \mathrm{and} \quad \mathcal{B}_{j \neq i} = \mathcal{E}^{(B)}.
\end{equation}
For a fixed input the problem of CPF reduces to a multi-ary quantum state discrimination \cite{helstrom1969quantum,cheflesQSD}.


\subsection{CPF using classical light}\label{CPFclass}

In a classical strategy, the source is described as a state with positive P-representation with a tensor product structure over the system, $\otimes_{i=1}^m \sigma$. This represents a statistical mixture of coherent states where each of the $m$ boxes is irradiated by $M$ modes and $M N_S$ photons in total. For such a source, the minimum error probability is lower-bounded by the Helstrom limit, which is the performance of the minimum error probability quantum receiver, given by
\begin{equation}
\mathrm{P}_{H,\mathrm{LB}} = \frac{m-1}{2m} F^4\left( \sigma^{(T)}, \sigma^{(B)} \right),   
\label{fidLB}
\end{equation}
where $F\left( \sigma^{(T)}, \sigma^{(B)}\right)$ is the fidelity between the states $\sigma^{(B/T)}=\mathcal{E}^{(B/T)}(\sigma)$, defined by ~\cite{weedbrook2012gaussian,barnum2002reversing}
\begin{equation}
\begin{split}
    F\left( \sigma^{(T)}, \sigma^{(B)}\right)&:= \left \Vert \sqrt{\sigma^{(T)}} \sqrt{\sigma^{(B)}} \right \Vert_1 \\
    &= \Tr \sqrt{\sqrt{\sigma^{(T)}}\sigma^{(B)}\sqrt{\sigma^{(T)}}}.
    \end{split}
\end{equation}

Suppose the target and background channels have transmissivity/gain $\mu_T$, $\mu_B$ and noise $E_T$, $E_B$, respectively. Then the general lower bound to the error probability for the classical benchmark, assuming a total of $m M$ modes and $m M N_S$ mean photons are irradiated over the entire system, is given by~\cite{zhuang2020entanglement}
\begin{equation}
\mathrm{P}_{H,\mathrm{LB}} = \frac{m-1}{2m} c^{2M}_{E_B,E_T} \exp \left[- \frac{2M N_S(\sqrt{\mu_B}-\sqrt{\mu_T})^2}{1 + E_B + E_T}  \right],
\label{class1}
\end{equation}
with
\begin{equation}
    c_{E_B,E_T} \equiv \left[ 1 + (\sqrt{E_B(1+(E_T))} - \sqrt{E_T(1+E_B)})^2 \right]^{-1}. 
    \label{class2}
\end{equation}

\subsection{CPF using entangled light}

For a quantum strategy, consider entangling each input state with ancillary idler to form a two-mode squeezed vacuum state, $\psi_{SI}:= \sum_{k=0}^{\infty} \sqrt{N_S^k/(N_S+1)^{k+1}} \ket{k,k}$, for each of the $m$ boxes. The idler systems are sent directly to the measurement with only the signals, each comprising $N_S$ photons per mode, probing the individual boxes. Thus, with $M$ modes per box, our entangled source takes the tensor product form $\psi_{SI}^{\otimes m M}$ which is acted on by the global quantum channel
\begin{equation}
\mathcal{E}_i \otimes \mathcal{I} = \left[\otimes_{k\neq i} (\mathcal{E}^{(B)}_k \otimes \mathcal{I}_k ) \otimes ( \mathcal{E}^{(T)}_i \otimes \mathcal{I}_i )\right],
\end{equation}
where $\mathcal{I}$ is the identity channel. This leads to an upper bound (UB) to the Helstrom limit given by~\cite{zhuang2020entanglement}
\begin{equation}
\mathrm{P}_{H,\mathrm{UB}} = (m-1)F^2 \left( \Xi^{(T)}, \Xi^{(B)} \right),
\label{QUB}
\end{equation}
where $\Xi^{(T/B)} = ( \mathcal{E}^{(T/B)} \otimes \mathcal{I} ) \psi_{SI}^{\otimes M}$.

Employing the generalised conditional-nulling (CN) receiver (see Ref.~\cite{zhuang2020entanglement} for full details) allows far better detection results to be achieved. Suppose there exist two partially unambiguous POVMs, one corresponding to the target channel, $\{\Pi_T^t,\Pi_T^b\}$, and another for the background channel $\{\Pi_B^b,\Pi_B^t\}$, acting such that
\begin{equation}
\Tr \left[ \Pi_T^t \Xi^{(T)} \right] = \Tr \left[ \Pi_B^b \Xi^{(B)} \right] =1.
\end{equation}
Entailing feed-forward across the entire channel arrangement and conditional-nulling of subsequent hypotheses, the CN receiver yields an overall error probability given by
\begin{equation}
\mathrm{P}_{\mathrm{err},m}^{\mathrm{CN}}(\zeta_1,\zeta_2) = \frac{1}{m} \frac{\zeta_2}{\zeta_1} \left( m \zeta_1 + (1-\zeta_1)^m -1 \right).
\label{CNerr}
\end{equation}
Here $\zeta_1 = \Tr \left[ \Pi_T^t \Xi^{(B)} \right] = p(H_1|H_0)$ and $\zeta_2 = \Tr \left[ \Pi_B^b \Xi^{(T)} \right] = p(H_0|H_1)$ are the Type-I and Type-II error probabilities associated with the two POVMs, respectively.%

Note that the CN receiver applies to QI case only in the limit of $N_S \ll 1$. In this limit one can construct some projector via the sum-frequency-generation (SFG) mechanism~\cite{FFSFG} to have $\Tr[\Xi^{(B)}]\simeq 1$ to leading order. Such a projector is possible because, in this limit, the idler is close to vacuum and, by allowing this approximation, $\Pi_B^b$ may always be different from the identity.

\section{Quantum target ranging}\label{secQTR}

\begin{figure}[t]
\resizebox{\linewidth}{!}{\input{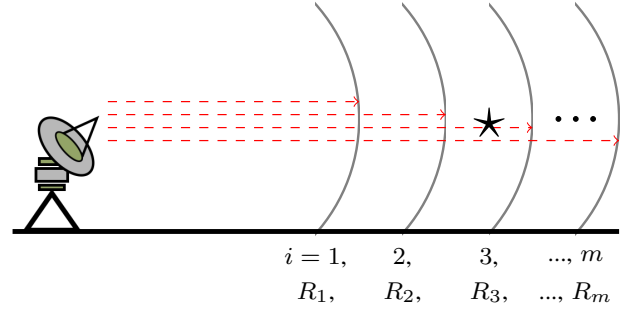}}
\caption{Schematic diagram of quantum target ranging (QTR). A range of surveillance from $R_{\mathrm{min}}$ to $R_{\mathrm{max}}$ is split into $m$ spherical shells with radius equal to the corresponding range $R_i$ for $i=1,\dots, m$. A total of $M$ signal-idler pulses with $N_S$ mean photons per mode are generated at well defined frequencies corresponding to each of the $m$ range bins, chosen such that only the radiation returned from that range bin is collected by its associated receiver. Each range bin has an associated signal-return time such that, upon its potential return, it may be recombined with the corresponding, retained idlers for joint measurement.}
\label{ranging}
\end{figure}

Quantum-enhanced CPF forms a crucial component towards enabling QI-based detection protocols to perform target-based metrology for measuring parameters such as location, range and speed. CPF can be viewed as a proxy for these problems in which we reformulate the protocol such that the $m$ hypotheses instead correspond to multiple space-time-frequency bins. Crucial to this is ensuring that the frequencies of each range bin's set of $M$ signal-idler pulses are chosen to ensure that their respective receivers only collect radiation returned from their associated range bins.  In such a setting, space bins would allow for the determination of target location, time bins would allow for range estimation, and frequency bins may be used for the measurement of speed via Doppler shift. The ranging problem is a major barrier towards real-world implementation of QI for target detection since the signal and idler modes must be synchronised at the receiver in order to perform an optimal joint measurement and obtain the best possible performances. For a target located an some unknown range to be determined, the return time is unknown and is thus problematic. 

A potential solution would involve a modification to the direction finding protocol (see Ref.~\cite{zhuang2020entanglement}), based on CPF, whereby we consider a fixed number $m$ of non-overlapping range bins across some range interval of interest, as shown in Fig.~\ref{ranging}. Then, by generating a sequence of $M$ signal-idler pulses, at a well-defined frequency, for each of the $m$ bins, the signals can be recombined (if returned at the corresponding range) with their respective idlers. All of the $m$ sectors are simultaneously probed with the total energy irradiated equal to $m M N_S$, where $N_S$ is the average number of photons per signal/idler mode and $M$ is the total number of modes (experiment repeats). In such a scenario, the total energy the target is exposed to is at most $m M N_S$, with this upper limit applying when the target is located in the final, $m$th, sector. 

Consider the thermal-loss channel $\mathcal{L}_{\mu}^{N}$ with loss parameter $\mu$ and mean number of thermal photons $N$ so its output noise is given  by $E=(1-\mu)N$. If the target is present in some sector, a proportion $\eta$ of the $N_S$ signal photons in each mode will be returned, which includes contributions from target reflectivity and transmission losses due to beam spreading. These will be mixed with thermal noise comprising $N_B$ photons per mode at the receiver and the $M$ signal modes are acted on by the target channel $\mathcal{E}^{(T)} = \left(\mathcal{L}^{N_B/(1-\eta)}_{\eta}\right)^{\otimes M}$ In contrast, if the target is absent, all of the $M$ signal modes are lost and the return consists only of background noise, with $N_B$ photons per mode. The corresponding background channel is given by $\mathcal{E}^{(B)} = \left(\mathcal{L}^{N_B}_{0}\right)^{\otimes M}$.

Using a quantum source comprising a tensor product of two-mode squeezed vacuum states $\psi_{SI}^{\otimes m M}$, each $M$-mode probing of each of the $m$ sectors yields an ensemble of outputs corresponding to the background and target channels, given by
\begin{equation}
    \Xi^{(T)} = \left[(\mathcal{L}^{N_B/(1-\eta)}_{\eta} \otimes \mathcal{I}) \psi_{SI} \right] ^{\otimes M},
\end{equation}
\begin{equation}
    \Xi^{(B)} = \left[(\mathcal{L}^{N_B}_{0} \otimes \mathcal{I}) \psi_{SI} \right] ^{\otimes M}.
\end{equation}
Computing the error probability, i.e., Eq.~(\ref{QUB}), for these states yields the following asymptotic bound for QTR,
\begin{equation}
    \mathrm{P}^{\mathrm{QTR}}_{H,\mathrm{UB}} \simeq (m-1) \exp \left( - \frac{M \eta N_S}{N_B+1}\right),
    \label{QRUB}
\end{equation}
in the limits of $N_S \ll 1$ and $M \gg 1$, keeping total energy $M N_S$ fixed.

By adapting the SFG receiver for QI~\cite{FFSFG} to the CN approach, the upper bound for QTR [Eq.~(\ref{QRUB})] may be significantly improved upon~\cite{zhuang2020entanglement}. After multiple SFG cycles and the application of differing two-mode squeezing operations, the photon-counting statistics of $\Xi^{(T/B)}$ allow for the realisation of two partially unambiguous POVMs for the CN receiver with corresponding mean error probability
\begin{equation}
    \mathrm{P}_{\mathrm{CN}}^{\mathrm{QTR}} \simeq \frac{1}{2} (m-1) \exp \left(- \frac{2 M \eta N_S}{N_B} \right).
    \label{QRCN}
\end{equation}

Consider now probing this set of $m$ possible range bins using a classical source. Described in Section \ref{CPFclass}, it comprises $M$ modes, each with $N_S$ photons, for each of the $m$ sectors, employing a total of $m M N_S$ photons across the entire protocol. For generic channel finding, the lower bound achievable using such a source is, using Eqs.~(\ref{class1}) and~(\ref{class2}) with $E_T = E_B = N_B$ and $\mu_T = \eta$, $\mu_B=0$, given by
\begin{equation}
    \mathrm{P}_{H,\mathrm{LB}} = \frac{m-1}{2m} \exp \left[- \frac{2 M \eta N_S}{2 N_B+1} \right].
    \label{classctrinit}
\end{equation}

However, the separation of each of the $m$ returning signals, while crucial for the QI-based protocol, is unnecessary for the classical source (its need arises from the fact that in any QI-based protocol, synchronised recombination is necessary for the joint-measurement upon which it is based). This has particular implications for target-ranging since regardless of the target's true location, any signal sent out will certainly return. It is only when using an entangled source where all but one (corresponding to the target) of the returns are essentially discarded. As a result there is, physically, no reason why the classical measurement process at the receiver should not incorporate the entire signal sent, that is, all of the $m M N_S$ coherent photons. In other words, the coherent pulse can be sent out in just one go to explore all the possible bins and the receiver just needs to be open and check for the return of the energetic pulse in each one of the bins. 
Thus, for classical target ranging (CTR), the lower bound of Eq.~(\ref{classctrinit}) becomes
\begin{equation}
    \mathrm{P}_{H,\mathrm{LB}}^{\mathrm{CTR}} = \frac{m-1}{2m} \exp \left[- \frac{2 m M \eta N_S}{2 N_B+1} \right].
    \label{classctr}
\end{equation}

To establish a quantum advantage in target-ranging, all that is needed is find a suitable parameter regime whereby the following relation (sufficient condition) is satisfied:
\begin{equation}
    \mathrm{P}_{\mathrm{CN}}^{\mathrm{QTR}} \leq \mathrm{P}_{H,\mathrm{LB}}^{\mathrm{CTR}}.
\end{equation}
Using Eqs.~(\ref{QRCN}) and~(\ref{classctr}), this is satisfied when 
\begin{equation}
    \ln m \leq 2 M \gamma \frac{N_B (2-m) +1}{2 N_B +1},
    \label{condition}
\end{equation}
where $\gamma = \eta N_S/N_B$ is the single-use signal-to-noise ratio (SNR).  For any value of $N_B>1$, the condition [Eq.~(\ref{condition})] cannot physically be satisfied for any $m>2$. Thus, it is not possible to prove that the QI-based ranging protocol can achieve a quantum advantage over its equivalent classical counterpart (at least with the current mathematical tools). This result is a  mathematical limitation of the approach since, while it provides a sufficient condition for there to exist a quantum advantage, it does not provide a necessary one.

While Eq.~\ref{condition} provides a sufficient condition for a quantum advantage in target ranging using the scheme proposed here, the following will outline necessary condition and its associated violations. Consider an adaptation of the CTR protocol (yielding the lower bound Eq.~(\ref{classctr})) to a scheme where one sends a coherent state pulse of average photon number $m M N_S$ to interrogate the $m$-bin range uncertainty interval. Performing homodyne detection alongside match filtering to maximise the SNR on each of the range bins' return signals yielding corresponding outputs $r_n$ where $1\leq n \leq m$. The receiver then chooses for the target to be present in the bin $m^{\star}=\mathrm{argmax}(r_n)$ since all hypotheses are equally likely. In the case of two bins, the minimum error probability for such a receiver is equivalent to the binary discrimination error given by

\begin{equation}
    \mathrm{P}^{\mathrm{CTR}\star}_{2} = Q \left(\sqrt{\frac{m M \eta N_S}{2N_B+1}} \right),
    \label{binary}
\end{equation}
where $Q(x)$ is defined as
\begin{equation}
    Q\left(x \right) = \sqrt{\frac{2}{\pi}}\int_{x}^{\infty} dy \exp \left(- \frac{y^2}{2} \right) \leq \frac{1}{2} \exp \left(- \frac{x^2}{2} \right),
     \label{binarybound}
\end{equation}
providing an exponentially tight upper bound. Extending to an $m$-array problem comprising $m-1$ background bins, the upper bound to the error probability for such a receiver, using Eqs.~(\ref{binary}) and~(\ref{binarybound}), is given by
\begin{equation}
    \mathrm{P}^{\mathrm{CTR}\star}_{\mathrm{UB}} \leq \frac{m-1}{2} \exp \left[ - \frac{m M \eta N_S}{2(2 N_B + 1)} \right].
    \label{classctrnew}
\end{equation}

Comparing this to Eq.~(\ref{QRCN}) for QTR with a CN receiver, our condition for a quantum advantage becomes
\begin{equation}
    m \leq 8 + \frac{4}{N_B}, 
    \label{conditionnewCN}
\end{equation}
which is violated for $m>8$ when $N_B>4$. Using a CN receiver, currently the best known tool for testing multiple quantum hypotheses, a quantum advantage may only potentially be realised for $m\leq8$.

\section{Conclusion}

In this paper we have outlined and studied a potential QI-based quantum ranging protocol based on multiple quantum hypothesis testing and channel discrimination, using recent results of quantum-enhanced CPF. By modeling a discrete set of ranges as multiple quantum channels, a quantum ranging protocol may be established by the distribution of signals and recombination with idlers across fixed time intervals. The very nature of the QI protocol demands that such an approach must be taken, since it is crucially dependent on the ability to recombine bosonic modes at the correct time in order to perform a joint measurement to harness any potential advantage. In the optimal classical scenario using a coherent state source, this energetic distribution across range bins (or, equivalently, time bins) is unnecessary and, for all practical purposes, would not be done. As such, the results of this paper show that, under fair energetic considerations and the currently available mathematical tools, it is not possible to show a quantum advantage in target ranging using a QI-based approach to CPF. 

\textbf{Acknowledgments.}~This work has been funded by the European Union's Horizon
2020 Research and Innovation Action under grant
agreement No. 862644 (FET-Open project: Quantum readout techniques
and technologies, QUARTET). A.K. acknowledges sponsorship by EPSRC Award No. 1949572 and Leonardo UK. The authors would like to thank Q. Zhuang for discussions and clarifications. The authors would also like to thank anonymous Reviewer 1 who suggested the possibility of determining a necessary condition for a quantum advantage using a CN receiver.


\begin{thebibliography}{99}                                                        
\bibitem{pirandola2018advances} S. Pirandola, B. R. Bardhan, T. Gehring, C. Weedbrook and S. Lloyd, \textit{Advances in photonic quantum sensing}, Nat. Photon. \textbf{12}, 724--733 (2018).

\bibitem {lloyd2008enhanced}S. Lloyd, \textit{Enhanced sensitivity of
photodetection via quantum illumination}, Science \textbf{321}, 1463-1465 (2008).

\bibitem {tan2008quantum}S.-H. Tan\textit{ et al.}, \textit{Quantum
illumination with Gaussian states}, Phys. Rev. Lett. \textbf{101}, 253601 (2008).

\bibitem{karsa2020generic} A. Karsa, G. Spedalieri, Q. Zhuang, and S. Pirandola, \textit{Quantum illumination with a generic Gaussian source}, Phys. Rev. Research \textbf{2}, 023414 (2020).

\bibitem{zhangexp}Z. Zhang, S. Mouradian, F. N. C. Wong and J. H. Shapiro, \textit{Entanglement-enhanced sensing in a lossy and noisy environment}, Phys. Rev. Lett. \textbf{114}, 110506 (2015).

\bibitem{lopaevaexp}E. D. Lopaeva, I. Ruo Berchera, I. P. Degiovanni, S. Olivares, G. Brida and M. Genovese, \textit{Experimental realization of quantum illumination}, Phys. Rev. Lett. \textbf{110}, 153603 (2013).

\bibitem{zhangent}Z. Zhang, M. Tengner, T. Zhong, F. N. C. Wong, and J. H. Shapiro, \textit{Entanglement’s benefit survives an entanglement-breaking channel}, Phys. Rev. Lett. \textbf{111}, 010501 (2013).

\bibitem {barzanjehmicrowave}S. Barzanjeh, S. Guha, C. Weedbrook, D. Vitali,
J. H. Shapiro, S. Pirandola, \textit{Microwave quantum illumination}, Phys.
Rev. Lett. \textbf{114}, 080503 (2015).

\bibitem{sacchi2005entanglement} M. F. Sacchi, \textit{Entanglement can enhance the distinguishability of entanglement-breaking channels}, Phys. Rev. A \textbf{72}, 014305 (2005).

\bibitem{shabirQI}S. Barzanjeh, S. Pirandola, D. Vitali, and J. M. Fink \textit{Microwave quantum illumination using a digital receiver}, Science Advances \textbf{6}, eabb0451 (2020).

\bibitem{luong2019receiver} D. Luong, C. W. Chang, A. M. Vadiraj, A. Damini, C. M. Wilson and B. Balaji, \textit{Receiver operating characteristics for a prototype quantum two-mode squeezing radar},  IEEE Transactions on Aerospace and Electronic Systems (2020). To appear.

\bibitem{brandsema2018current} M. Brandsema, \textit{Current readiness for quantum radar implementation}, 2018 IEEE Conference on Antenna Measurements \& Applications (CAMA), 1 (2018).

\bibitem{zhuang2020entanglement} Q. Zhuang and S. Pirandola, \textit{Entanglement-enhanced testing of multiple quantum hypotheses}, Comm. Phys. \textbf{3}, 103 (2020).


\bibitem {helstrom1969quantum}C. W. Helstrom, \textit{Quantum detection and
estimation theory}, J. of Stat. Phys \textbf{1}, 231 (1969).

\bibitem{cheflesQSD} A. Chefles, \textit{Quantum state discrimination}, Contemp. Phys. \textbf{41}, 401 (2000).

\bibitem{barnettQSD}S. M. Barnett and S. Croke, \textit{Quantum state discrimination}, Adv. Opt. Photon. \textbf{1}, 238 (2009).


\bibitem{hayashi2017quantum} M. Hayashi, \textit{Quantum Information Theory: Mathematical Foundation.} (Springer-Verlag, Berlin, 2017).

\bibitem{zhuang2020b}Q. Zhuang and S. Pirandola, \textit{Ultimate limits for multiple quantum channel discrimination}, Phys. Rev. Lett. \textbf{125}, 080505 (2020).



\bibitem{weedbrook2012gaussian} C. Weedbrook, S. Pirandola, R. Garc{\'\i}a-Patr{\'o}n, N. J. Cerf, T. C. Ralph, J. H. Shapiro, and S. Lloyd, \textit{Gaussian quantum information}, Rev. Mod. Phys. \textbf{84}, 621 (2012

\bibitem{barnum2002reversing} H. Barnum and E. Knill, \textit{Reversing quantum dynamics with near-optimal quantum and classical fidelity}, J. Math. Phys. \textbf{43}, 2097(2002).

\bibitem {FFSFG}Q. Zhuang, Z. Zhang, and J. H. Shapiro, \textit{Optimum
mixed-state discrimination for noisy entanglement-enhanced sensing}, Phys.
Rev. Lett. \textbf{118}, 040801 (2017).





\end{thebibliography}
\end{document}